\documentclass[acmtog]{acmart}

\acmSubmissionID{1077}

\title{Beyond Positional Encoding: A 5D Spatio-Directional Hash Encoding}

\author{Philippe Weier}
\email{weier@cg.uni-saarland.de}
\orcid{0000-0003-4276-1804}
\affiliation{%
    \institution{Meta}
    \country{Switzerland}
}
\affiliation{%
    \institution{Saarland University}
    \country{Germany}
}

\author{Lukas Bode}
\email{lbode@meta.com}
\orcid{0000-0002-8710-8561}
\affiliation{%
    \institution{Meta}
    \country{Switzerland}
}

\author{Philipp Slusallek}
\email{slusallek@cg.uni-saarland.de}
\orcid{0000-0002-2189-2429}
\affiliation{%
    \institution{Saarland University}
    \country{Germany}
}

\author{Adrián Jarabo}
\email{ajarabo@meta.com}
\orcid{0000-0001-9000-0466}
\affiliation{%
    \institution{Meta}
    \country{Spain}
}

\author{Sébastien Speierer}
\email{speierers@meta.com}
\orcid{0000-0001-6919-7567}
\affiliation{%
    \institution{Meta}
    \country{Switzerland}
}

\usepackage{tcolorbox}
\usepackage{minted}
\definecolor{hlcol}{RGB}{255,200,200}

\newcommand{\mintedpseudocode}[1]{
    \begin{tcolorbox}[arc=1pt,colback=black!3!white,colframe=black!16!white,boxrule=0.3mm,before skip=1mm,after skip=1mm,top=.5mm,bottom=.5mm,left=1mm,right=1mm]
         \inputminted[escapeinside=||,mathescape=true,fontsize=\footnotesize,linenos]{python}{#1}
    \end{tcolorbox}
}

\newfloat{algorithm}{thp}{lop}
\floatname{algorithm}{Algorithm}

\usepackage{stfloats}
\usepackage{lipsum} 
\usepackage{wrapfig}
\usepackage{colortbl}
\usepackage[capitalize,nameinlink]{cleveref}

\usepackage{tabularx}
\usepackage{mathtools}
\usepackage{soul}

\newcommand{\TODO}[1]{ {\color{red}\textbf{TODO : }#1}  }

\definecolor{RevisionColor}{RGB}{2,95,217}

\definecolor{PhilippeColor}{RGB}{2,95,217}
\definecolor{SebastienColor}{RGB}{170,46,200}
\definecolor{LukasColor}{RGB}{20,180,22}
\definecolor{AdrianColor}{RGB}{210,20,22}

\newcommand{\pw}[1]{\textcolor{PhilippeColor}{[P: #1]}}
\newcommand{\seb}[1]{\textcolor{SebastienColor}{[S: #1]}}
\newcommand{\lukas}[1]{{\textcolor{LukasColor}{[L: #1]}}}
\newcommand{\adrian}[1]{{\textcolor{AdrianColor}{[A: #1]}}}

\newcommand{\pwDone}[1]{\pw{\sout{#1}}}
\newcommand{\sebDone}[1]{\elie{\sout{#1}}}
\newcommand{\lukasDone}[1]{\tamy{\sout{#1}}}
\newcommand{\adrianDone}[1]{\alex{\sout{ #1 }}}

\renewcommand{\TODO}[1]{}
 
 \renewcommand{\pw}[1]{}
 \renewcommand{\seb}[1]{}
 \renewcommand{\lukas}[1]{}
 \renewcommand{\adrian}[1]{}
 \renewcommand{\pwDone}[1]{}
 \renewcommand{\sebDone}[1]{}
 \renewcommand{\lukasDone}[1]{}
 \renewcommand{\adrianDone}[1]{}

\usepackage{xspace}

\newcommand{\hashsphere}{\emph{hash-sphere}}
\newcommand{\hashgridsphere}{\emph{hash-grid-sphere}}
\newcommand{\hashgrid}{\emph{hash-grid}}

\newcommand{\discretize}[1]{\text{discr}(#1)}

\begin{document}


\begin{teaserfigure}
    \centering
    \includegraphics{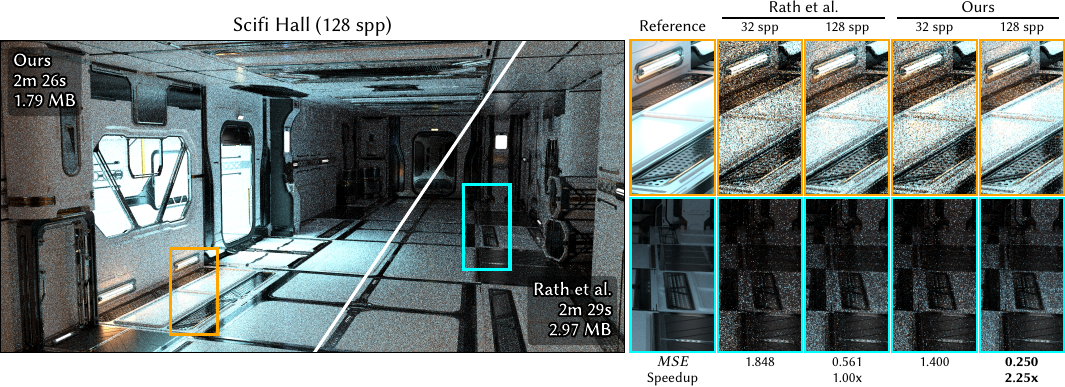}%
    \caption{We use our five-dimensional \hashgridsphere\ encoding to compactly learn the radiance distribution on a scene in the context of neural path guiding~\citep{rath_neural_resampling}. For an equal rendering time, our encoding learns a more accurate spatio-directional incident radiance distribution, drastically improving variance in scenes with complex global illumination, compared with Rath's \hashgrid\ with one-blob encoding~\citep{Mueller22}. }
    \label{fig:teaser}%
\end{teaserfigure}


\begin{abstract}
In this work, we propose a new spatio-directional neural encoding that is compact and efficient, and supports all-frequency signals in both space and direction. Current learnable encodings focus on Cartesian orthonormal spaces, which have been shown to be useful for representing high-frequency signals in the spatial domain. However, directly applying these encodings in the directional domain results in distortions, singularities, and discontinuities. As a result, most related works have used more traditional encodings for the directional domain, which lack the expressivity of learnable neural encodings. We address this by proposing a new angular encoding that generalizes the hash-grid approach from~\citet{Mueller22} to the directional domain by encoding directions using a hierarchical geodesic grid. Each vertex in the geodesic grid stores a learnable latent parameter, which is used to feed a neural network. Armed with this directional encoding, we propose a five-dimensional encoding for spatio-directional signals. We demonstrate that both encodings significantly outperform other hash-based alternatives. We apply our five-dimensional encoding in the context of neural path guiding, outperforming the state of the art by up to a factor of 2 in terms of variance reduction for the same number of samples.
\end{abstract}



\maketitle

\section{Introduction}
Directional functions are omnipresent in computer graphics and in particular in light transport simulation, where its basic unit, radiance, is defined in the directional domain. This makes the representation of spherical signals a cornerstone of modern physics-based rendering. For that reason, several representations with different properties have been proposed for modeling functions such as environment or cached probe lighting, data-driven scattering functions, visibility distributions, or incident radiance distributions for path guiding. 

On the other hand, neural encodings have been demonstrated to be enormously useful when modeling spatial functions, drastically improving the learning and reconstruction of such three-dimensional Cartesian signals \citep{mildenhall2020nerf,sitzmann2020implicit,Mueller22}. However, the power of these encodings have been mostly neglected in the angular domain, where traditional representations (e.g., spherical harmonics or spherical Gaussians) are usually preferred. This limits the functional representative power for complex all-frequency directional signals, resulting in severe approximations or ad-hoc solutions. 

In this work, we propose a new directional encoding that inherently captures all-frequency signals in a compact, efficient way. It is based on the \hashgrid\ representation proposed for the spatial domain by \citet{Mueller22}, which we transfer to the directional domain. Our directional encoding, which we call \hashsphere, is based on a hierarchical recursive geodesic grid, which we use to index a hash table of trained features. These are used as input for a small multilayered perceptron (MLP) that returns the directional value.
We then combine this directional representation with M\"uller's spatial \hashgrid\, creating a five-dimensional spatio-directional encoding (the \hashgridsphere) which is able to compactly represent complex high-frequency spatially-varying and view-dependent functions such as appearance. 

Our \hashsphere\ and \hashgridsphere\ encodings are drop-in replacements for directional or spatio-directional encodings respectively. We implement them in Dr.Jit and Mitsuba~\cite{Mitsuba3,Jakob2022DrJit}, and demonstrate their applicability in the context of neural path guiding in scenes with complex incident radiance configurations (\cref{fig:teaser}), where we obtain a $2.2\times$ increase in performance compared to the state-of-the-art~\citep{rath_neural_resampling}. 

In summary, our contributions are:
\begin{itemize}
    \item The \hashsphere, an efficient, compact all-frequency encoding for directional signals,
    \item the \hashgridsphere, a 5D neural encoding that combines directional and spatial encoding, 
    \item and a prototype path guiding application demonstrating the applicability of our 5D spatio-directional encoding. 
\end{itemize}



\section{Related work}
\paragraph*{Spherical Basis}
The most common spherical basis used in graphics are spherical harmonics (SH), which have been used for modeling scattering functions~\citep{westin1992predicting}, radiance transfer functions and environment maps~\citep{ramamoorthi2001efficient, sloan2002precomputed}, and emission in radiance fields~\citep{Kerbl23}: They are analogous to the Fourier transform in the spherical domain, but unfortunately they scale poorly with high-frequency signals. %
Wavelets~\citep{ng2003all,ng2004triple} and spherical wavelets~\citep{schroder1995spherical} on the other hand, are better suited at representing all-frequency signals, but at the price of losing continuity of the represented signal, discretized in the different mother wavelets. 
Finally, radial basis functions represent the signal via a mixture model of basis (e.g., von Mishes-Fisher distributions~\citep{fisher1953dispersion,han2007frequency} and (anisotropic) spherical Gaussians~\citep{xu2013anisotropic}) are able to represent a continuous all-frequency signal, but are limited to the number of predefined modes, scale poorly with multimode high-frequency signals, their fitting is challenging as soon as their dimensionality increases, and are difficult to interpolate except for the trivial case of a single basis. 
In contrast, our \hashsphere\ is able to represent a continuous all-frequency signal compactly, and can be combined natively with an efficient spatial encoding.

\paragraph*{Neural and Learnable Encodings}

Recent advances in neural scene representations and generative models have been fueled by a number of efficient spatial encodings, helping neural networks to learn spatial information. These encodings include explicit~\citep{mildenhall2020nerf} and implicit~\citep{sitzmann2020implicit} Fourier encoding of the position, adaptive network capacity based on data spatial complexity~\citep{takikawa2022variable}, or compact neural projections from 3D to sets of two-dimensional planes~\citep{chan2022efficient}. Other works proposed to use smaller MLPs by using explicit spatial encoding, e.g., via spatial subdivision structures to represent varying detail~\citep{martel2021acorn}. In a similar spirit, but focusing on efficient training and evaluation, \citet{Mueller22} used a compact learnable hash table to store the latent information at discrete positions corners in a hierarchy of grids. These are later used to feed an output MLP, allowing orders of magnitude faster learning and evaluation. Our method builds on top of this work, by extending it to directional and spatio-directional signals. None of these works target the representation of directional signals, which are usually represented using spherical harmonics or the one-blob encoding \citep{muller2019neural}.


\paragraph*{Path Guiding}
We demonstrate our \hashgridsphere\ by applying it to learn the local incident radiance distribution in the context of path guiding~\cite{vorba2019path}. Most previous works have combined a spatial-subdivision structure with an explicit angular encoding, including adaptive histograms~\cite{muller2017practical}, mixtures of radial functions including Gaussians~\cite{vorba2014line,herholz2016product}, von Mishes-Fisher distributions~\cite{ruppert2020robust}, cosine lobes~\cite{bashford2012significance} or linearly-transformed cosines~\cite{diolatzis2020practical}. \citet{dodik2022path} proposed using a spatio-directional Gaussian mixture of the incident radiance. \citet{muller2019neural} learned a neural sampleable representation based on normalizing flows, but resulted in a very costly training and evaluation. Closer to us, \citet{rath_neural_resampling} proposed using the \hashgrid\ combined with the one-blob encoding to model the incident radiance, which they show how to sample using resampled importance sampling (RIS)~\cite{talbot2005importance}. We apply Rath's methodology using our new representation, which allows for a more faithful representation of the angular domain, translating to better guiding especially in areas with complex directional radiance signal. 



\begin{figure}
    \centering
    \includegraphics[width=\linewidth]{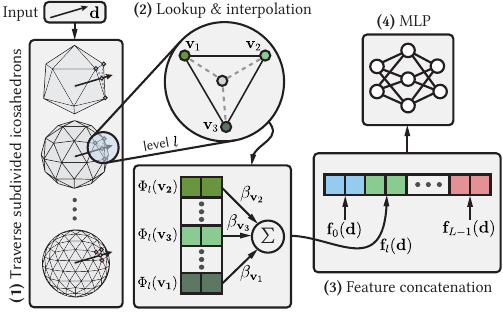}
    \caption{\textbf{The \hashsphere\ encoding}. For an input direction $\mathbf{d}$: \textbf{(1)} We traverse a hierarchy of $L$ recursively-subdivided icosahedral grids defined in the unit sphere, identifying the enclosing triangle at each level. \textbf{(2)} At each level $l$, we retrieve learnable features  $\theta_l[ \Phi_l(\mathbf{v}) ]$ from the triangle's three vertices (using direct indexing for coarse levels, and hashing for fine levels) and interpolate them using barycentric coordinates $\boldsymbol{\beta}_l$. \textbf{(3)} Features from all levels are concatenated into $\mathbf{f}(\mathbf{d})$. \textbf{(4)} A small MLP maps the concatenated feature vector $\mathbf{f}(\mathbf{d})$ to the final output.\TODO{change figure to use L-1 rather than N and ideally bold font etc. to match math} \adrian{in 2, h(v) is not defined. It should be $\theta_l[\phi_l(v)]$ according to Eq.2. Or at least $\phi_il(v)$. Also, missing beta, and how it is used for interpolation. }
    }
    \label{fig:directional_encoding}
\end{figure}

\section{Overview}
\label{sec:method_overview}

We propose a hierarchical feature encoding that efficiently represents high-frequency signals on the sphere $\mathbb{S}^2$ and on the product space of positions and directions $\mathbb{R}^3 \times \mathbb{S}^2$. Our approach is motivated by the observation that existing neural encodings, while highly effective for spatial signals, fail to properly handle the spherical topology of the directional domain. Previous approaches either map directions to Cartesian coordinates (creating a sub-optimal space for directional signals) or to polar coordinates (introducing singularities and distortions at the poles).

At the core of our method is the \hashsphere, a learnable directional encoding based on a recursive geodesic grid (\cref{fig:directional_encoding}). Unlike latitude-longitude parameterizations, the geodesic tessellation provides a near-uniform discretization of the sphere while avoiding polar singularities. We describe this encoding in \cref{subsec:spherical_hash} and demonstrate its effectiveness by learning HDR environment maps, where it outperforms hash-grid alternatives that suffer from parametric distortions (\cref{fig:directional_encoding_analysis}).

Building on the \hashsphere, we introduce the \hashgridsphere, a joint encoding for the five-dimensional product space $\mathbb{R}^3 \times \mathbb{S}^2$. This encoding couples the spatial hash-grid of \citet{Mueller22} with our hierarchical geodesic grid, enabling compact representation of spatially-varying, view-dependent functions such as radiance (\cref{fig:5d_encoding_illustration}). Critically, the \hashgridsphere\ performs interpolation in a geometrically meaningful way in both the spatial and directional domains, allowing it to generalize to novel viewpoints, a capability that eludes naive extensions of the hash-grid to higher dimensions. We present this encoding in \cref{subsec:joint_encoding} and evaluate it on a sparse-view radiance field reconstruction task (\cref{fig:5d_encoding_plots,fig:5d_encoding_images}).

Finally, we demonstrate the practical utility of our encodings by applying the \hashgridsphere\ to neural path guiding (\cref{sec:guiding}), where learning the spatially-varying incident radiance distribution benefits directly from our encoding's ability to capture high-frequency directional signals.

\section{The \hashsphere: Spherical hash encoding}
\label{subsec:spherical_hash}

\begin{figure*}
    \centering
    \includegraphics[width=0.9\linewidth]{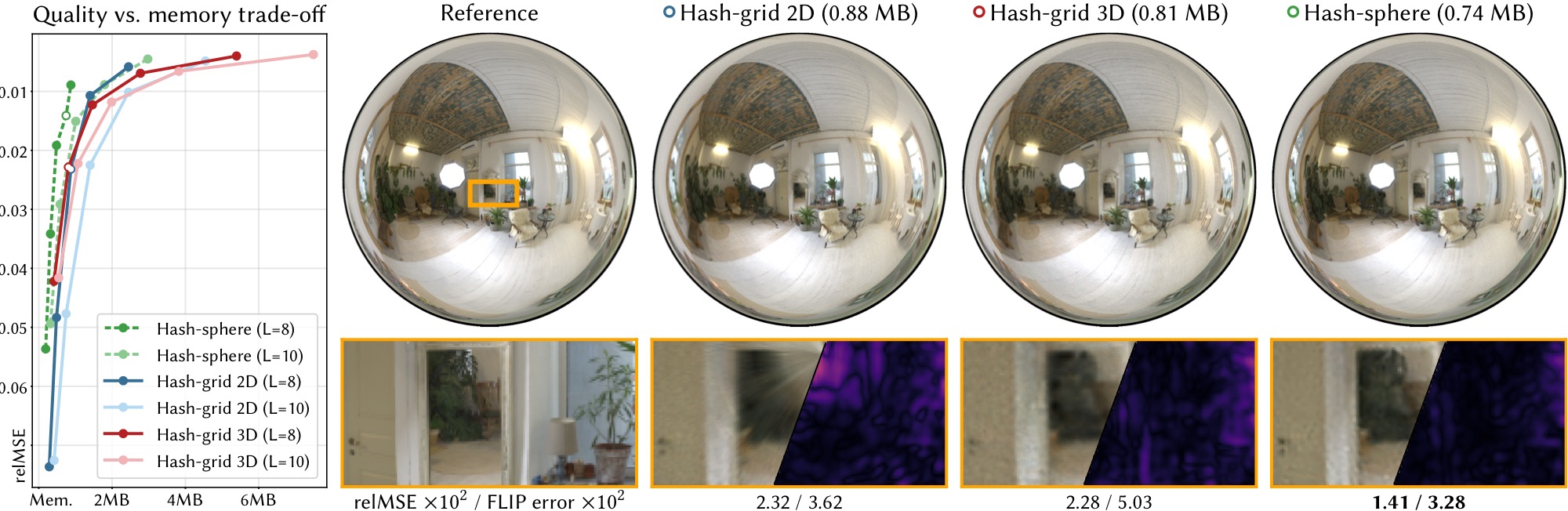}
    \caption{Quality vs. memory trade-off representing an HDR environment map. We compare our \hashsphere\ against 2D and 3D hash-grid variants. The 2D hash-grid (polar parameterization) achieves comparable quality at mid-latitudes but suffers from severe distortions near the poles (inset). The 3D hash-grid (Cartesian) avoids polar artifacts but introduces interpolation-related artifacts due to working on a sub-optimal space for directional signals. 
    Our \hashsphere\ provides consistent angular resolution across the sphere with an intuitive relationship between subdivision levels and frequency content. Memory includes both encoding parameters and MLP weights. Additional analyses with other environment maps can be found in the supplemental material. }
    \label{fig:directional_encoding_analysis}
\end{figure*}

\paragraph{Description}
We first introduce a standalone directional encoding defined on the unit sphere $\mathbf{d} \in \mathbb{S}^2$. To avoid the polar singularities of latitude-longitude grids and the low-frequency limitation of spherical harmonics, we utilize a recursive geodesic grid to provide a uniform discretization of the directional domain.
The encoding operates over $L$ resolution levels. At level $l=0$, the sphere is tessellated into the 20 faces of a regular icosahedron. At each subsequent level $l > 0$, every triangle from level $l-1$ is subdivided into four sub-triangles, with new vertices reprojected onto the sphere. For each triangle vertex we store a learnable parameter $\theta$ which is used to compute an input feature vector $\mathbf{f}_l(\mathbf{d})$ per level, ultimately feeding an MLP that produces the final encoded directional function.

   More specifically, for an input direction $\mathbf{d}$, we traverse the multilevel hierarchy (\cref{fig:directional_encoding}.1). At each level $l$, we identify the enclosing triangle with vertices $V_{\mathbf{d}, l} = \{\mathbf{v}_1, \mathbf{v}_2, \mathbf{v}_3\}$ and compute the barycentric coordinates $\boldsymbol{\beta}_l = \{\beta_{\mathbf{v}_1}, \beta_{\mathbf{v}_2}, \beta_{\mathbf{v}_3}\}$ (\cref{fig:directional_encoding}.2) (details in the supplementary material). The feature vector $\mathbf{f}_l(\mathbf{d})$ is computed by linearly interpolating learnable parameters $\theta$ stored at these vertices (\cref{fig:directional_encoding}.2).
To maintain a bounded memory footprint as the number of vertices $|V_l|$ increases, we employ a hybrid indexing scheme similar to the one proposed by \citet{Mueller22}. For coarser levels where the total vertex count $|V_l|$ is small, we use a unique dense mapping. For finer levels where $|V_l|$ exceeds a fixed maximum capacity per level $T$, we resolve vertex indices using a hash function $h_\text{sphere}(\cdot)$. The output feature vector $\mathbf{f}_l(\mathbf{d})$ at level $l$ is thus given by
\begin{equation}
    \mathbf{f}_l(\mathbf{d}) = \sum_{\mathbf{v} \in V_{\mathbf{d}, l}} \beta_\mathbf{v} \cdot \theta_l [ \Phi_l(\mathbf{v}) ],
\end{equation}
where $\theta_l$ is the hash table of size $|\theta_l| = \min(T,|V_l|)$ storing the per-vertex learnable latent parameters, and $\Phi_l(\mathbf{v})$ is the level-specific indexing function:
\begin{equation}
    \Phi_l(\mathbf{v}) =
    \begin{cases}
    \text{idx}(\mathbf{v}) & \text{if } |V_l| \leq T, \\
    h_\text{sphere}(\mathbf{v}) \mod T & \text{otherwise}.
    \end{cases}
\end{equation}
Here, $\text{idx}(\mathbf{v})$ is the unique index of the vertex in the subdivided icosahedron. The hash function $h_\text{sphere}$ maps discretized integer coordinates of the vertex vectors to table indices following
\begin{equation}
    h_\text{sphere}(\mathbf{v}) = \bigoplus_{j = \{x,y,z\}} \discretize{v_j}\cdot\pi_j,
    \label{eq:hashing_sphere}
\end{equation}
where $\oplus$ denotes the bitwise XOR operation, $\pi_j$ are large prime numbers, and $\discretize{v}=\lfloor(1+v)\cdot\gamma\rfloor$ discretizes the floating-point coordinate $v$, with $\gamma$ a large integer.
Note that we hash the directional vertices in Cartesian coordinates; while polar coordinates could be used, these would introduce distortions, singularities and discontinuities in the collision pattern.
After computing the features for level $l$, the directional triangle $V_{\mathbf{d}, l}$ is refined by checking which of its four sub-triangles contains $\mathbf{d}$, determining $V_{\mathbf{d}, l+1}$ for the next iteration. The final directional encoding concatenates features from all levels into a single vector $\mathbf{f}(\mathbf{d}) = [\mathbf{f}_0(\mathbf{d}); \mathbf{f}_1(\mathbf{d}); \ldots; \mathbf{f}_{L-1}(\mathbf{d})]$, which is then passed to a shallow MLP whose size and output are application-dependent (\cref{fig:directional_encoding}.4). \Cref{code:our_hashsphere} summarizes the complete encoding procedure; helper functions are detailed in the supplemental material.

\begin{algorithm}[t]
    \mintedpseudocode{code/hashsphere_concise.py}
    \caption[Hash-sphere pseudocode]{
        \label{code:our_hashsphere} Pseudocode for our \hashsphere\ encoding. See our supplemental material for the specific implementations of \texttt{icosahedron\_intersection} and \texttt{refine\_triangle}.}
\end{algorithm}

\paragraph{Analysis}

To evaluate the effectiveness of our \hashsphere\ encoding, we compare it against other learnable hash-based encodings on the task of compressing HDR environment maps. Given a ground-truth environment map, we optimize each encoding to reconstruct the radiance for randomly sampled directions on the unit sphere. In all cases we use a small MLP with 2 hidden layers and 16 neurons per layer, identity activations, and an exponential output activation. For training, we use a relative $L_2$ loss, and optimize for 512 steps using Adam with a learning rate of $0.01$.

\Cref{fig:directional_encoding_analysis} shows the quality-versus-memory trade-off for three encodings: our \hashsphere, a 2D hash-grid with directions mapped to polar coordinates, and a 3D hash-grid with directions encoded in Cartesian coordinates. For all encodings, we use 2 features per level, while hash-grid variants use a base resolution of 8 with a per-level scaling factor of 2. We vary the per-level hash table maximum size $T$ from $2^{14}$ to $2^{18}$, and test both 8 and 10 levels to span a range of memory budgets. Note that the storage of coarse levels depends on the number of vertices stored, which explains the small differences in storage for each method even for the same $T$.

While a well-configured 2D hash-grid can achieve similar quality-versus-memory trade-offs at low latitudes, it suffers from severe distortions around the poles due to the non-uniform Jacobian of the spherical parameterization. The 3D hash-grid avoids these polar artifacts by operating in Cartesian space, but representing the 2D manifold in 3D results in discontinuities between voxels on the same level and between levels, which impacts the interpolation and thus the reconstruction, and introduces significant overhead (30\%, see \cref{tab:encoding_benchmark}).
%
%
In contrast, our \hashsphere\ works directly on the 2D manifold of the spherical surface, which avoids the problem of both previous Cartesian encodings, introducing minimal overhead (4\% compared to the polar encoding). It additionally provides a direct relationship between the number of subdivision levels and angular resolution: Each additional level quadruples the number of triangular faces, providing consistent frequency content across the entire sphere. We do not compare against spherical harmonics, as they require prohibitively many coefficients to represent high-frequency signals and our focus is on learnable feature encodings.

\section{The \hashgridsphere: Joint spatio-directional encoding}
\label{subsec:joint_encoding}

\begin{figure}
    \centering
    \includegraphics[width=\linewidth]{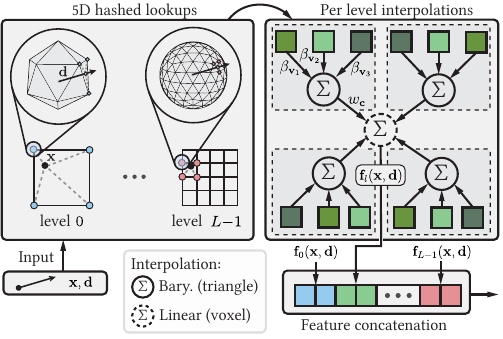}
    \caption{\textbf{The \hashgridsphere\ encoding.} For an input position-direction pair $(\mathbf{x}, \mathbf{d})$: At each level $l$, we locate both the enclosing spatial voxel (with 8 corners $C_{\mathbf{x},l}$) and the enclosing triangle in the sphere (with 3 vertices $V_{\mathbf{d},l}$). We retrieve learnable parameters for all $8 \times 3 = 24$ corner-vertex pairs (12 shown here), using either direct indexing or the joint hash function $h_{\text{joint}}$ depending on the grid size. Parameters are interpolated using the product of trilinear weights $w_\mathbf{c}$ and barycentric coordinates $\beta_\mathbf{v}$. Features from all levels are then concatenated and, typically, passed to an MLP. Compared to the \hashsphere\ (\cref{fig:directional_encoding}), the key addition is the coupling of spatial and directional grids through joint indexing, enabling compact representation of 5D spatio-directional signals.\TODO{change figure to use L-1 rather than N and ideally bold font etc. to match math, also add dashed boxes around each sum to represent a voxel}}
    \label{fig:5d_encoding_illustration}
\end{figure}

\paragraph{Description}
To capture spatially-varying functions (e.g., radiance), we extend the \hashsphere\ concept to the product space $\mathbb{R}^3 \times \mathbb{S}^2$. At each level $l$, we maintain the geodesic grid at subdivision depth $l$ along with a spatial voxel grid of resolution $N_l^3$.

For an input query $(\mathbf{x}, \mathbf{d})$ with $\mathbf{x}\in\mathbb{R}^3$, the encoding proceeds hierarchically. At level $l$, we locate the spatial voxel corners $C_{\mathbf{x}, l}$ and the directional triangle vertices $V_{\mathbf{d}, l}$. In the case of the spatial domain, we get the trilinear weights for each corner $\mathbf{w}_l=\{w_{\mathbf{c}_{1:8}}\}$. The joint feature vector for level $l$ is obtained by interpolating parameters mapped from the coupled coordinates, following
\begin{equation}
    \mathbf{f}_l(\mathbf{x}, \mathbf{d}) = \sum_{\mathbf{c} \in C_{\mathbf{x}, l}} \sum_{\mathbf{v} \in V_{\mathbf{d}, l}} \left( w_{\mathbf{c}} \cdot \beta_{\mathbf{v}} \right) \cdot \theta_l [ \Phi_l(\mathbf{c}, \mathbf{v}) ],
\end{equation}
which are later concatenated to build $ \mathbf{f}(\mathbf{x}, \mathbf{d})=[\mathbf{f}_0(\mathbf{x}, \mathbf{d});...;\mathbf{f}_{L-1}(\mathbf{x}, \mathbf{d})]$. As in \cref{subsec:spherical_hash}, we employ a hybrid indexing scheme: For coarse levels where the grid is small, we use direct indexing; for finer levels, we use the hash function $h_{\text{joint}}$ defined as
\begin{equation}
    \Phi_l(\mathbf{c}, \mathbf{v}) =
    \begin{cases}
    \text{idx}(\mathbf{c}, \mathbf{v}) & \text{if } |C_l| \cdot |V_l| \leq T, \\
    h_{\text{joint}}(\mathbf{c}, \mathbf{v}) \mod T & \text{otherwise},
    \end{cases}
\end{equation}
where $h_{\text{joint}}$ extends $h_\text{sphere}$ by combining the spatial voxel corner $\mathbf{c}$ with the directional vertex $\mathbf{v}$ as
\begin{equation}
    h_{\text{joint}}(\mathbf{c}, \mathbf{v}) = \left( \bigoplus_{i=\{x,y,z\}} c_i \cdot \pi_{\mathbf{x},i} \right) \oplus \left( \bigoplus_{j=\{x,y,z\}} \discretize{v_j} \cdot \pi_{\mathbf{d},j} \right),
\end{equation}
with 
$\pi_{\mathbf{x},i}$ and $\pi_{\mathbf{d},j}$ being a set of six large prime numbers. Note that the voxel corner coordinates are already integers in the range $[0, N_l)$. As in the \hashsphere, the size of the per-level parameters table is $|\theta_l| = \min(T, |C_l|\cdot|V_l|)$.
\Cref{fig:5d_encoding_illustration} illustrates the changes required for our joint 5D encoding with respect to the directional \hashsphere.

While the spatial grid resolution scales at a given factor (typically $2\times$) at each level, the directional icosahedron does not need to be refined at the same rate. We therefore define a mapping $m: \{0, \ldots, L-1\} \rightarrow \{0, \ldots, L_\mathbf{d}\}$ that specifies the directional subdivision level $l_\mathbf{d} = m(l)$ used at spatial level $l$. The directional triangle is refined whenever $m(l+1) > m(l)$. In our implementation, we use $m(l) = \lfloor l / 2 \rfloor$, which refines the directional grid every two spatial levels. In our experiments we found this to provide a good balance between spatial and angular resolution for typical view-dependent effects.
Overall, the \hashgridsphere\ enables compact representation of 5D spatio-directional signals with bounded memory
while maintaining meaningful interpolation in both domains. \Cref{code:our_hashgridsphere} summarizes the complete procedure.

\begin{algorithm}[t]
    \mintedpseudocode{code/hashgridsphere_concise.py}
    \caption[Hash-grid-sphere pseudocode]{
        \label{code:our_hashgridsphere} Pseudocode for our \hashgridsphere\ encoding.}
\end{algorithm}

\begin{figure}
    \centering
    \includegraphics[width=\linewidth]{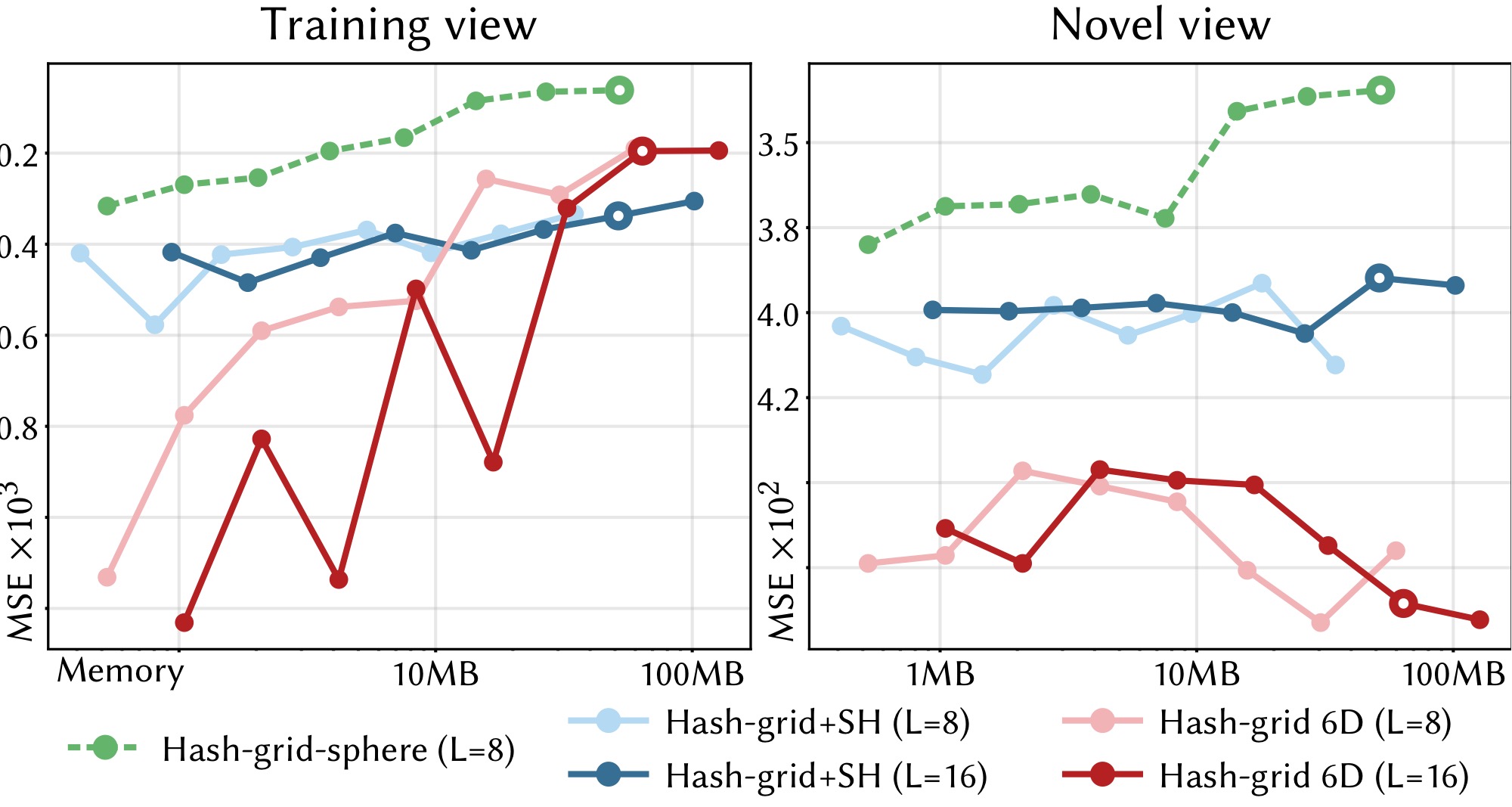}
    \caption{Error vs. memory for radiance field reconstruction on the \textsc{Phone} scene. We report reconstruction error for both training and novel views across three encodings. The 3D hash-grid + SH cannot capture high-frequency view dependence, producing blurred results. The 6D hash-grid overfits training views but fails on novel views due to ill-defined directional interpolation. Our \hashgridsphere\ achieves low error on both training and novel views, demonstrating meaningful generalization. Corresponding images for the highlighted configurations are shown in \cref{fig:5d_encoding_images}.}
    \label{fig:5d_encoding_plots}
\end{figure}

\paragraph{Analysis}
We evaluate the effectiveness of our \hashgridsphere\ encoding with a sparse-view radiance field-like reconstruction task. We place 256 cameras at a fixed distance from the center of a mesh and capture ground-truth images of a complex, view-dependent appearance. Assuming known surface hit positions, we train each encoding to predict the outgoing radiance as a function of the 3D surface position and 2D viewing direction---a 5D signal living on the surface of the mesh.

We compare three encodings: Our \hashgridsphere, a 3D hash-grid concatenated with degree-9 spherical harmonics (100 SH coefficients), and a 6D hash-grid treating the problem as a pure 6D spatial signal. For all hash-grid components, we use a base resolution of 16, a per-level scaling factor of 2, and 2 features per level. We vary the hash table maximum size $T$ from $2^{13}$ to $2^{21}$ to span different memory budgets, and test 8 and 16 levels $L$ for the baseline encodings. For our \hashgridsphere, we use $L=8$ levels with 4 directional levels (refining every two spatial levels).
In all cases we use a small MLP with 2 hidden layers and 16 neurons
per layer, LeakyReLU activation, and Sigmoid output activation.
We optimize for 4096 iterations using Adam with a learning rate of $0.005$, which is sufficient for all methods to converge.
\Cref{fig:5d_encoding_plots} reports the reconstruction error versus memory footprint for both a training view and a held-out novel view. Results are shown in \cref{fig:5d_encoding_images} while animated results can be found in our supplementary video.

The results reveal fundamental differences in how each encoding handles the 5D signal. The 3D hash-grid + SH approach can represent low-frequency view dependence but over-blurs highlights as degree-9 spherical harmonics still lack the capacity to encode high-frequency directional variation. The 6D hash-grid has sufficient representational power to overfit the training views, but fails catastrophically on novel views, since the encoding does not respect the spherical topology, leading to meaningless view extrapolation. In addition, it requires $2^6=64$ hash table lookups per level and six-linear interpolation, resulting in a significant performance drop (over 90\%, see \cref{tab:encoding_benchmark}). 

Our \hashgridsphere\ is the only encoding that both reconstructs high-frequency signals accurately \emph{and} generalizes meaningfully to novel views. This is because interpolation in our geodesic grid is geometrically consistent over the directional domain. Furthermore, by decoupling the number of directional levels from spatial levels, we can control the directional resolution independently, avoiding overfitting in sparse-view settings while maintaining high spatial detail. This flexibility is a key advantage of our factored encoding design. Our encoding requires 3 times more hash table lookups than the hash-grid + SH encoding, making it over a 40\% more expensive \cref{tab:encoding_benchmark}), though it pays-off given the significant boost on accuracy.


\section{Application: Neural Path Guiding}
\label{sec:guiding}

\begin{table}[t]
    \centering
    \caption{Performance comparison of the various neural encodings used in our figures. We report timings in milliseconds for a joint forward/backward pass of a full HD frame ($\approx2.07M$ samples). All experiments where run on an NVIDIA GeForce RTX 4070, using half-float precision. The configurations in bold are the one used in \cref{fig:guiding_results,fig:guiding_nirc,fig:teaser}}
    \label{tab:encoding_benchmark}
    \begin{tabular}{l@{\hspace{0pt}}c@{\hspace{0pt}}c@{\hspace{0pt}}c@{\hspace{0pt}}c}
        \toprule
        \textbf{Encoding} & Size (MB) & Time (ms) & Speedup \\
        \midrule
        \multicolumn{4}{l}{\cref{fig:directional_encoding_analysis}: \textit{Directional}} \\
        \midrule
        Hash-sphere & 0.74 & $6.89 \pm 0.01$ & 0.96x \\
        Hash-grid 2D & 0.88 & $6.64 \pm 0.01$ & 1.00x \\
        Hash-grid 3D & 0.81 & $9.49 \pm 0.01$ & 0.70x \\
        \midrule
        \multicolumn{4}{l}{\cref{fig:5d_encoding_images}: \textit{Spatio-Directional - Radiance Field}} \\
        \midrule
        Hash-grid 3D + SH & 55.73 & $120.57 \pm 0.86$ & 1.00x \\
        Hash-grid 6D & 67.11 & $1392.92 \pm 4.08$ & 0.09x \\
        Hash-grid-sphere & 52.10 & $301.17 \pm 0.12$ & 0.40x \\
        \midrule
        \multicolumn{4}{l}{\cref{fig:teaser,fig:mlp_size_comparison,fig:guiding_nirc,fig:guiding_results}: \textit{Spatio-Directional - Path Guiding}} \\
        \midrule
        Hash-grid 3D + OB, small MLP & 2.92 & $15.51 \pm 0.01$ & 1.32x \\
        Hash-grid 3D + OB, large MLP & \textbf{2.97} & \textbf{20.42}$\pm$\textbf{0.02} & \textbf{1.00x} \\
        Hash-grid-sphere, small MLP & \textbf{1.80} & \textbf{34.47}$\pm$\textbf{0.02} & \textbf{0.59x} \\
        Hash-grid-sphere, large MLP & 1.83 & $45.36 \pm 0.08$ & 0.45x \\
        \bottomrule
    \end{tabular}
\end{table}

To demonstrate the practical benefits of our \hashgridsphere\ encoding, we apply it to neural path guiding, where learning the spatially-varying incident radiance distribution directly benefits from our encoding's ability to capture high-frequency directional signals. We build upon the recent work of \citet{rath_neural_resampling}, replacing their radiance encoding with our \hashgridsphere.

\paragraph{Background}
Rath et al.\ propose to learn a neural representation of the incident radiance field $\mathcal{N}_\Theta(\mathbf{x}, \mathbf{d})$ with trainable parameters $\Theta$, which predicts the incident radiance at position $\mathbf{x}$ from direction $\mathbf{d}$. Unlike methods that learn normalized, directly sampleable distributions~\citep{muller2019neural}, their approach learns an unnormalized radiance function and relies on resampled importance sampling (RIS)~\citep{talbot2005importance} to generate samples.
At each shading point, RIS first draws $M$ candidate directions from a source distribution $q(\mathbf{d})$ (typically combining BSDF sampling with a lightweight guiding distribution); then, one candidate is selected with probability proportional to a target weight $w(\mathbf{d}_i) = \hat{p}(\mathbf{d}_i) / q(\mathbf{d}_i)$, where $\hat{p}(\mathbf{d}_i) $ is the product of the learned incident radiance, the BSDF, and the path throughput. Increasing $M$ improves the approximation to the target distribution, at the cost of additional encoding and network evaluations per sample.

Rath's full system includes several additional components: A defensive resampling scheme, asynchronous GPU training while the CPU renders, and an optimized resampling candidate allocation (ORCA) framework that automatically determines spatially-varying candidate counts to maximize rendering efficiency. Their encoding concatenates a spatial hash-grid with the one-blob directional encoding~\citep{muller2019neural}, feeding the result to an MLP.

\paragraph{Our Setup}
We adopt the core RIS-based sampling framework of Rath et al., but focus specifically on evaluating the benefit of our \hashgridsphere\ encoding, replacing their hash-grid + one-blob encoding with our \hashgridsphere. To isolate the effect of the encoding, we use a \emph{constant} candidate count $M$ across all shading points (rather than ORCA's spatially-varying allocation), and we only use BSDF sampling as our source distribution. This allows for a direct comparison of encoding quality without confounding factors from adaptive allocation schemes.

For the baseline, we use Rath's recommended hash-grid configuration: 8 levels with base resolution 8, per-level scaling of 2, hash table max size $T=2^{16}$, and 4 features per level, concatenated with an 8-bin one-blob directional encoding. For our \hashgridsphere, we configure it to roughly match their memory footprint: 8 spatial levels with 4 directional subdivision levels, per-level scaling of 2, hash table max size $T=2^{16}$, and 2 features per level.
We optimize both methods using Adam with learning rate $1e^{-3}$, a relative $L_2$ loss and an $L_2$ regularization with $\lambda=1e^{-5}$, as recommended in their paper. We train both representations for 2048 iterations, which we found is sufficient for both methods to reach their respective optima and ensures we are comparing the representational capacity of the encodings rather than training dynamics.

\begin{figure*}
    \centering
    \includegraphics[width=\linewidth]{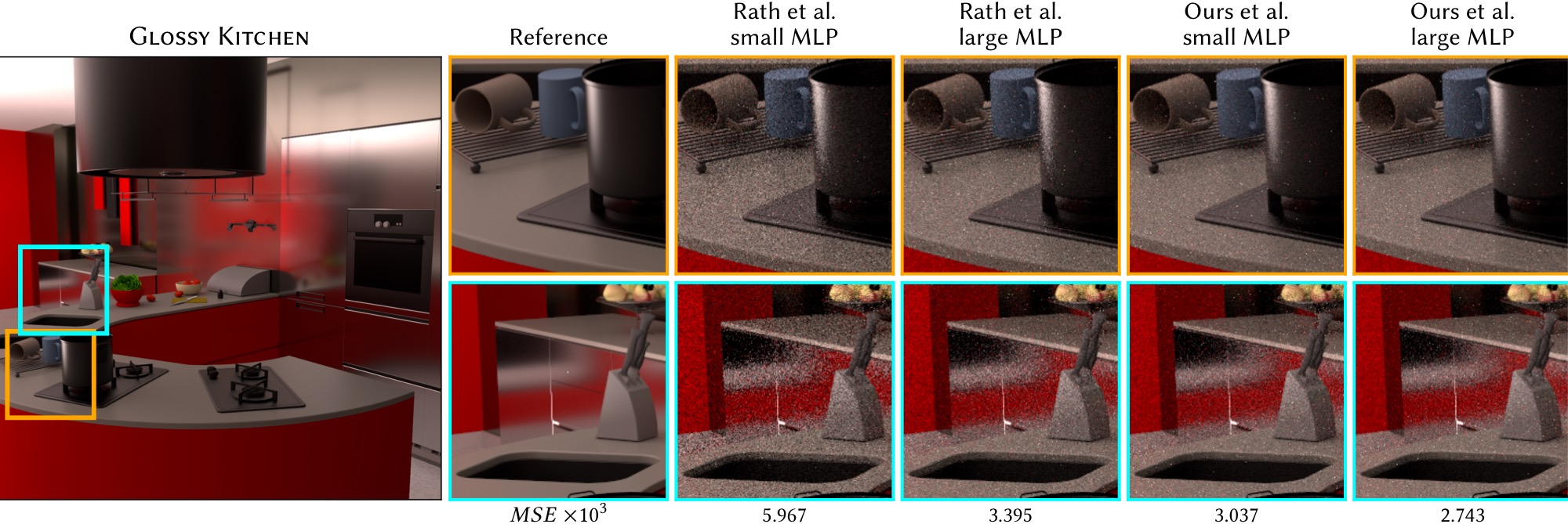}
    \caption{Effect of MLP capacity on guiding quality. We compare a small MLP (16 neurons, 2 hidden layers) against a larger MLP (64 neurons, 3 hidden layers) for both encodings with $M=16$ candidates. The baseline hash-grid + one-blob encoding requires a large MLP to capture locally varying directional distributions, as its directional component is global. Our \hashgridsphere\ achieves comparable quality with the small MLP, since local high-frequency variations are captured by the encoding itself. We use the small MLP for our method and the large MLP for the baseline in all subsequent experiments.}
    \label{fig:mlp_size_comparison}
\end{figure*}

\paragraph{Network Capacity Requirements}
\Cref{fig:mlp_size_comparison} compares the effect of MLP size on guiding quality for both encodings. We test a small MLP (16 neurons, 2 hidden layers) against the larger MLP recommended by Rath et al.\ (64 neurons, 3 hidden layers), with a fixed candidate count of $M=16$. In both cases, we use ReLU activations and an exponential output activation. For Rath's encoding, the larger MLP is essential: Since the positional and directional components are simply concatenated, the directional encoding is inherently global, and the network must learn to spatially modulate the directional response, requiring significant capacity to capture locally varying, high-frequency incident radiance distributions.

Our \hashgridsphere\ encoding, in contrast, does not require a large MLP to achieve good results. Because the directional and spatial components are jointly indexed, local high-frequency directional variations are captured directly by the encoding, leaving less work for the MLP. In fact, both the small and large MLP's performs comparably with our encoding. We therefore use the small MLP for all subsequent experiments with our method, while retaining the large MLP for the baseline as recommended by Rath et al.

\paragraph{Performance Considerations}
\Cref{tab:encoding_benchmark} reports the computational cost of each configuration. Our \hashgridsphere\ with a small MLP is about $40\%$ slower than the baseline with a large MLP. However, this modest slowdown is more than offset by the improved guiding quality. To enable equal-time comparisons, we allocate more candidates to the baseline method ($M=16$ for our approach, $M=24$ for Rath's) which approximately equalizes the per-sample computational cost. In this configuration, \cref{fig:teaser} shows that our encoding achieves a $2.25\times$ variance reduction for equal rendering time in scenes with complex radiance distributions. Note that this improvement should be even more noticeable in production scenes, where sampling paths become the main bottleneck.

\begin{figure*}
    \centering
    \includegraphics[width=\linewidth]{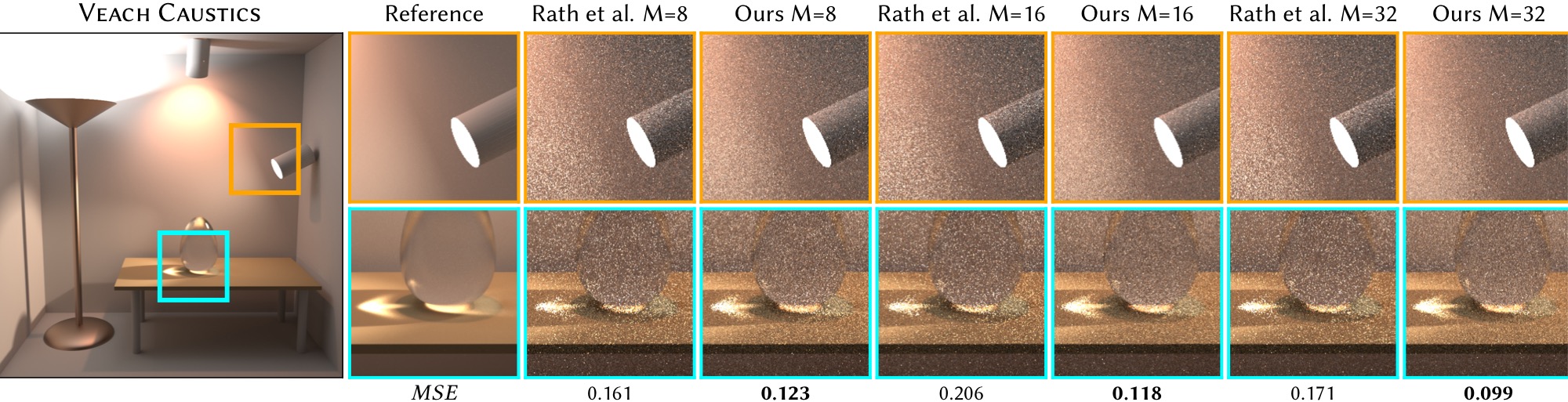}
    \includegraphics[width=\linewidth]{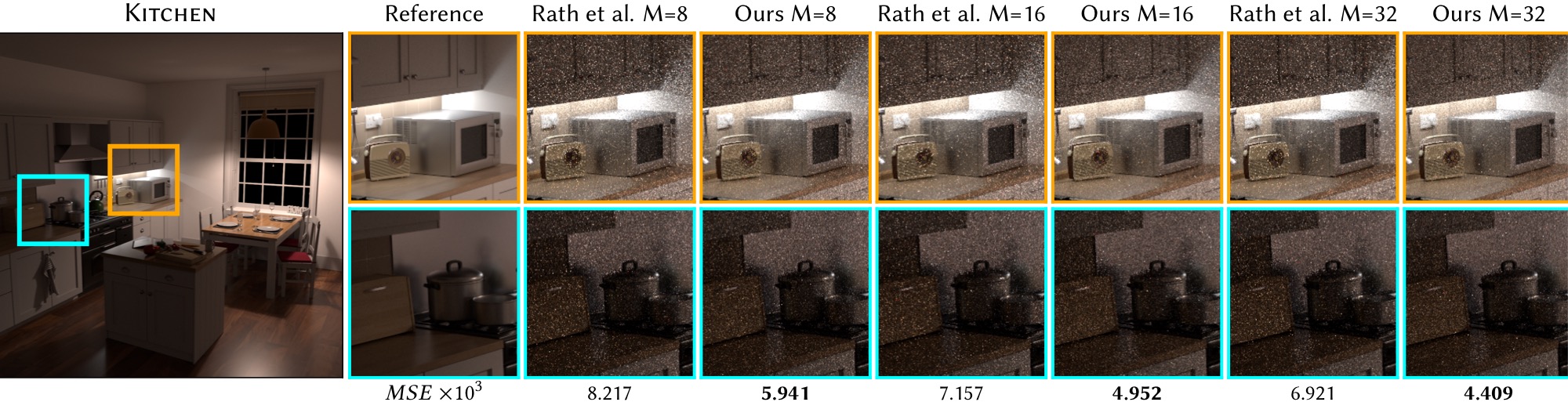}
    \includegraphics[width=\linewidth]{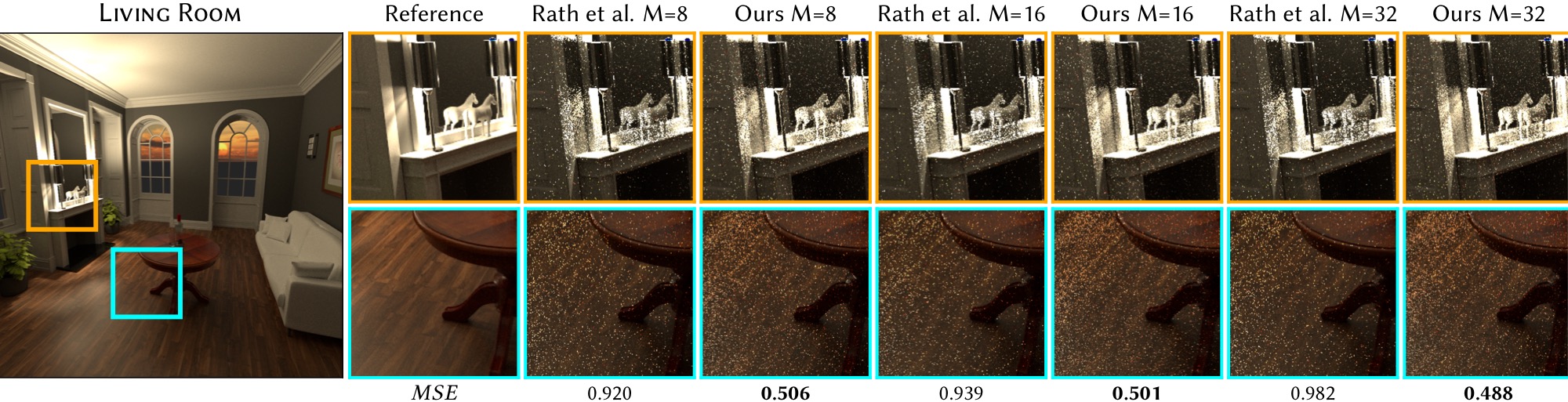}
    \includegraphics[width=\linewidth]{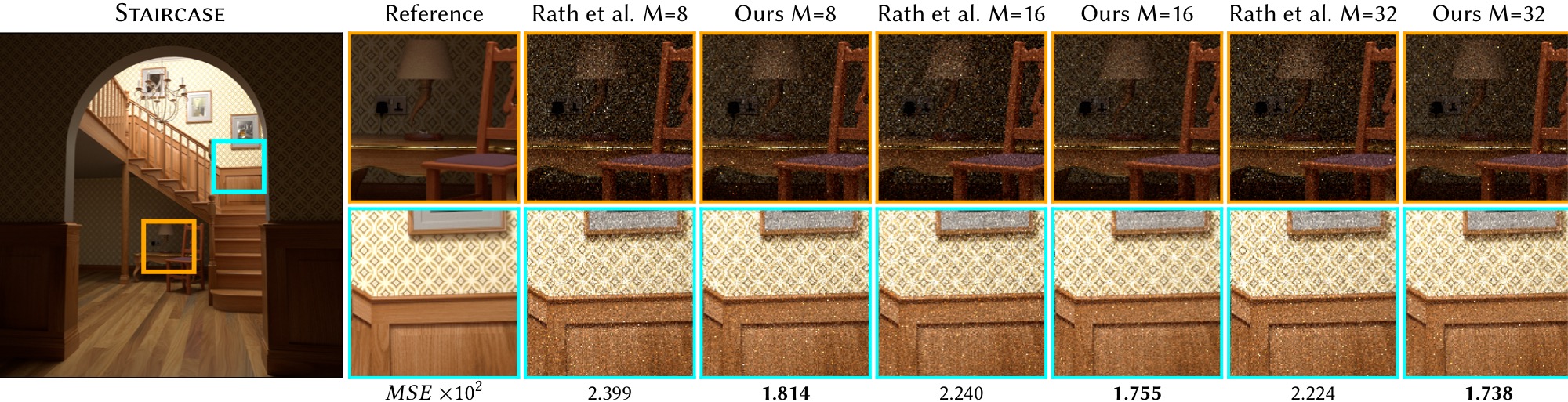}
    \caption{Path guiding comparison across four scenes with varying candidate counts $M \in \{8, 16, 32\}$. Both methods improve with more candidates, but our \hashgridsphere\ benefits more due to its higher-fidelity representation of the incident radiance distribution. Our method with $M=8$ consistently outperforms the baseline with $M=32$. The improvement is most pronounced in scenes with complex multi-modal lighting and high-frequency visibility changes.}
    \label{fig:guiding_results}
\end{figure*}

\paragraph{Guiding Quality Evaluation}
\Cref{fig:guiding_results} evaluates rendering quality across four scenes using both guiding representations with varying candidate counts ($M \in \{8, 16, 32\}$). Both methods benefit from increasing $M$, as more candidates allow RIS to better approximate the learned distribution. However, because our encoding captures  the incident radiance distribution more faithfully, the benefit of additional candidates is even more pronounced with our approach.
Notably, our method with only $M=8$ candidates consistently outperforms the baseline with $M=32$ candidates. This is particularly significant as Rath et al.\ report that their ORCA scheme can assign over 140 candidates to challenging regions; our encoding's improved fidelity suggests even greater relative benefits in such scenarios. 

The improvement is especially pronounced in scenes with complex multi-modal lighting, such as caustics from multiple light sources (\textsc{Veach Caustics}). The baseline struggles to capture sharp, localized illumination features with its global directional encoding, while our \hashgridsphere\ naturally adapts to such high-frequency signals. Even in predominantly diffuse scenes (\textsc{Staircase}), our encoding maintains an advantage by better capturing high-frequency visibility changes that the baseline encoding cannot resolve.

\begin{figure}
    \centering
    \includegraphics[width=\columnwidth]{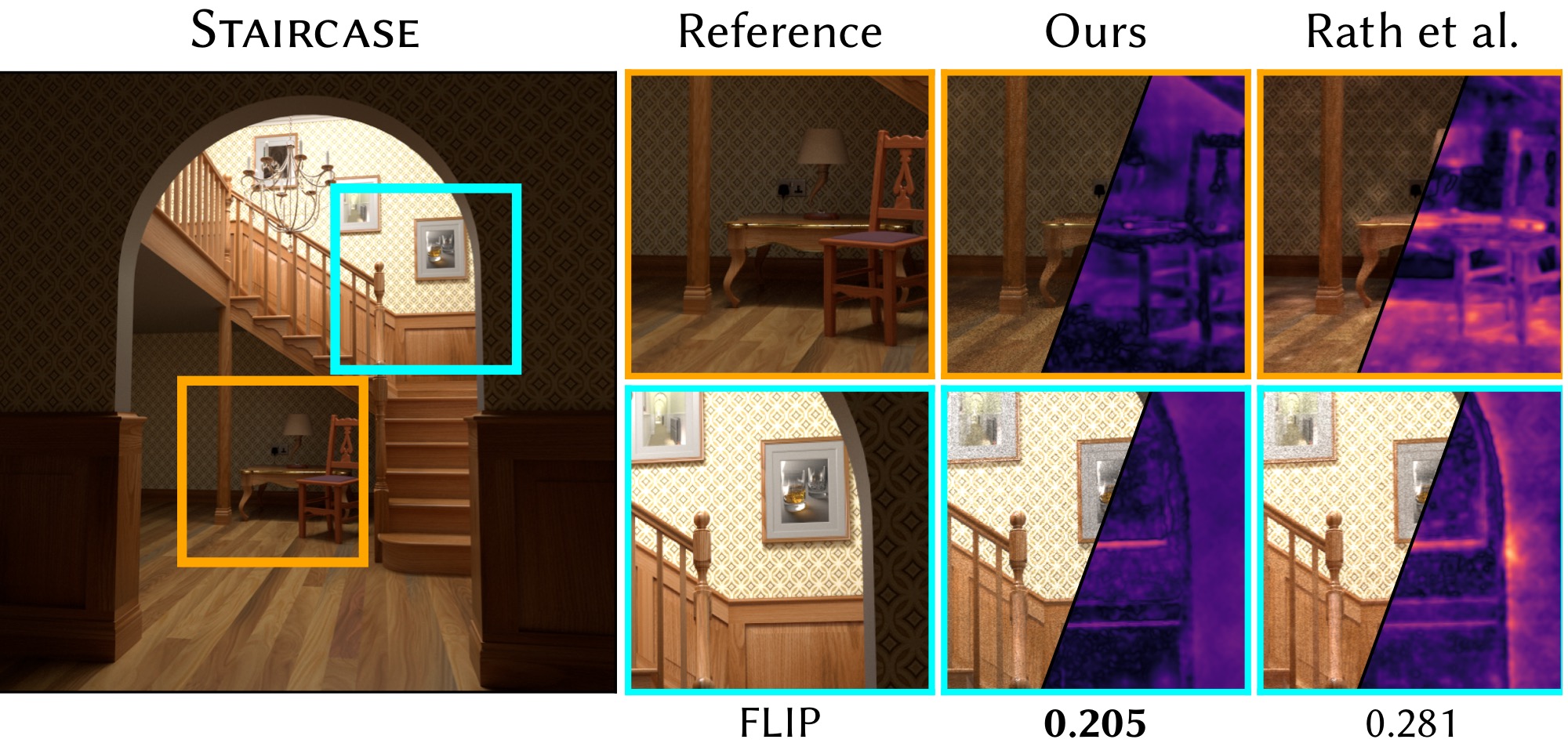}
    \caption{Equal sample neural incident radiance caching comparison. Both approaches perform similarly for simple diffuse indirect lighting (bottom). However, for complex indirect lighting with glossy materials (bottom), Rath et al. produce splotchy artifacts, while our encoding robustly handles high-frequency view-dependent indirect illumination.}
    \label{fig:guiding_nirc}
    \vspace{-0.5cm}
\end{figure}

\paragraph{Neural Incident Radiance Caching}
In addition to path guiding, \citet{rath_neural_resampling} show that their learned incident radiance representation can serve as a neural radiance cache, directly providing radiance estimates for indirect illumination. \Cref{fig:guiding_nirc} shows that our encoding also improves this application.
The scene compares our approach against the baseline in two different lighting scenarios: In the simpler case (bottom row)---diffuse materials with at most one indirect bounce to reach the light source---both encodings perform comparably. However, in the more challenging lighting (top row)---where light often requires two or more indirect bounces and a glossy material on the table creates view-dependent reflections---the baseline produces splotchy artifacts due to its inability to capture the complex spatio-directional radiance distribution. Our \hashgridsphere\ handles this case robustly by faithfully representing the highly-directional spatially-varying incident radiance.

\section{Discussion}

\paragraph{Limitations and Future Work}
A core limitation of our approach is the increased overhead with respect to the hash-grid plus a lightweight directional encoding (one-blob, SH). Our method requires three times more hash queries than the hash-grid alone, and depending on the complexity of the angular signal this might not pay off. However, we have shown that, as soon as the angular signal exhibits high frequencies, our method quickly outperforms the baseline, by requiring less levels and a smaller MLP. Investigating automatic techniques to adapt the number of icosahedron subdivision levels based on the frequency content of the signal might be an interesting future work, which would allow to provide further control over performance and quality.

An additional issue, inherited from the hash-grid approach, is that the current approach cannot perform level-of-detail in part because the hash map stores latent representations that eventually feed the MLP. Thus, in its current form, there is no explicit mapping between levels in the hierarchy and signal frequency.

\begin{figure*}
    \centering
    \includegraphics[width=\linewidth]{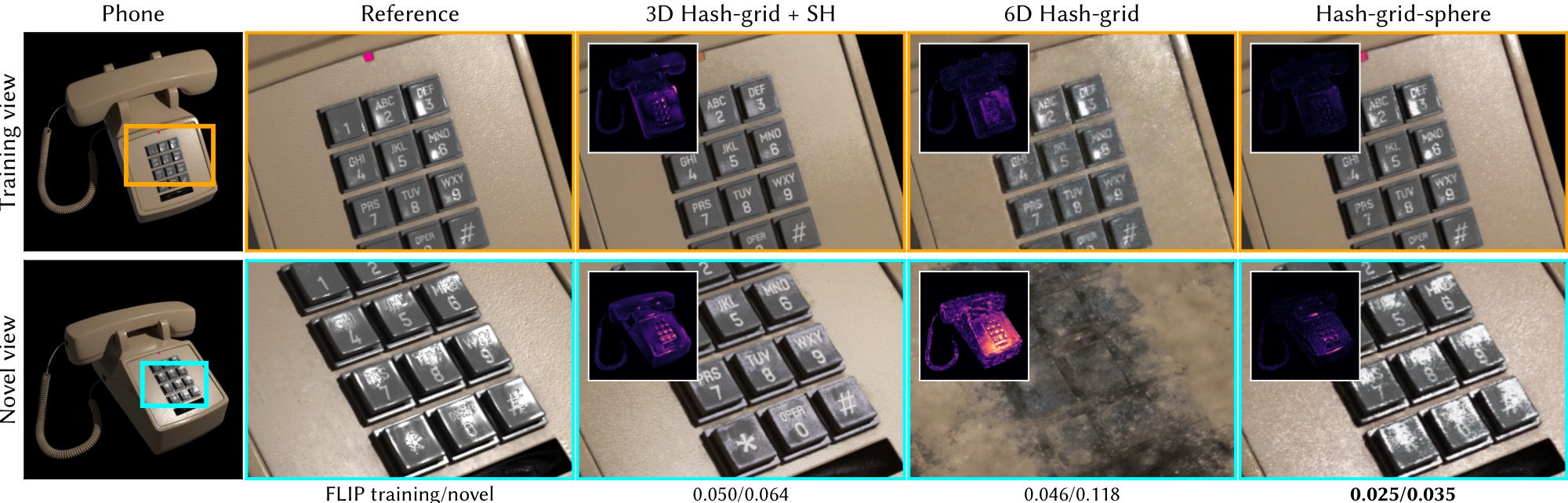}
    \caption{Qualitative comparison of sparse-view radiance field reconstruction. Top row: a training view; bottom row: a novel view. Concatenating a 3D hash-grid with spherical harmonics (SH) produces over-blurred results that lack high-frequency detail. The 6D hash-grid can overfit training views but cannot interpolate meaningfully in the directional domain, significantly degrading its performance in novel views. Our \hashgridsphere\ is the only encoding that reconstructs high-frequency signals while generalizing to novel viewpoints.}
    \label{fig:5d_encoding_images}
\end{figure*}

Finally, we have demonstrated and evaluated our work in a single application (neural path guiding), but show only hints of its potential modeling radiance fields (\cref{fig:5d_encoding_images}) and for incident radiance caching (\cref{fig:guiding_nirc}). Investigating how well our method works in these, and other domains (e.g., neural BSDFs \citep{sztrajman2021neural}) is a promising avenue for future work. 


\paragraph{Conclusion}
We have shown that our new directional and spatio-directional encodings are an efficient and compact tool to represent signals defined in the angular domain. They are extremely compact thanks to the hashing mechanism, and do not present the limitations of the current neural encodings designed to work on Cartesian spaces. Our encoding does not present singularities or discontinuities, has no area distortion on the unit sphere, and allows smooth spatio-directional interpolation even in high-frequency signals. It is, to our knowledge, the first neural encoding to directly represent 5D spatio-directional signals compactly, and we hope it will serve as a drop-in replacement from previous methods, at least in scenarios where the angular domain requires an additional level of complexity beyond low-frequency representations.


\begin{acks}

\end{acks}


\bibliographystyle{ACM-Reference-Format}
\bibliography{bibliography}

\newpage

\end{document}


\maketitle
This supplemental document provides additional implementation details and experimental results for our \hashsphere\ and \hashgridsphere\ encodings.
\section{Implementation Details}
\label{sec:implementation}
\subsection{Helper Functions}
\Cref{code:hashsphere_utils} provides pseudocode for the helper functions referenced in the main paper: \texttt{icosahedron\_intersection} finds the enclosing triangle on the base icosahedron and computes barycentric coordinates using the Möller–Trumbore algorithm~\cite{moller1997fast}, while \texttt{refine\_triangle} subdivides a triangle into four sub-triangles and identifies which one contains the query direction.

\begin{algorithm*}
    \mintedpseudocode{code/hashsphere_utils.py}
    \caption{Pseudocode for icosahedron traversal helpers. \texttt{icosahedron\_intersection} finds the enclosing triangle on the base icosahedron by comparing the query direction against face normals, then computes barycentric coordinates via ray-triangle intersection. \texttt{refine\_triangle} subdivides the current triangle into four sub-triangles by computing edge midpoints (projected onto the unit sphere) and determines which sub-triangle contains the direction $\mathbf{d}$.}
    \label{code:hashsphere_utils}
\end{algorithm*}

\subsection{Memory Efficiency}
The indexing function $\Phi_l(\mathbf{v})$ operates differently depending on the level:
\begin{itemize}
    \item \textbf{Dense levels} ($|V_l| \leq T$): We use unique vertex indices, requiring a precomputed lookup table that maps face indices to vertex indices. This table must be stored for each dense level.
    \item \textbf{Hashed levels} ($|V_l| > T$): We hash the discretized vertex position directly. Since vertex positions are computed on-the-fly during triangle refinement (as normalized midpoints of parent edges), no additional storage is needed beyond the parameter table $\theta_l$.
\end{itemize}
This means storage for the subdivision structure scales only with the number of dense levels, not the total number of levels $L$.
\section{Additional Results}
\label{sec:additional_results}
\begin{figure*}
    \centering
    \includegraphics[width=\linewidth]{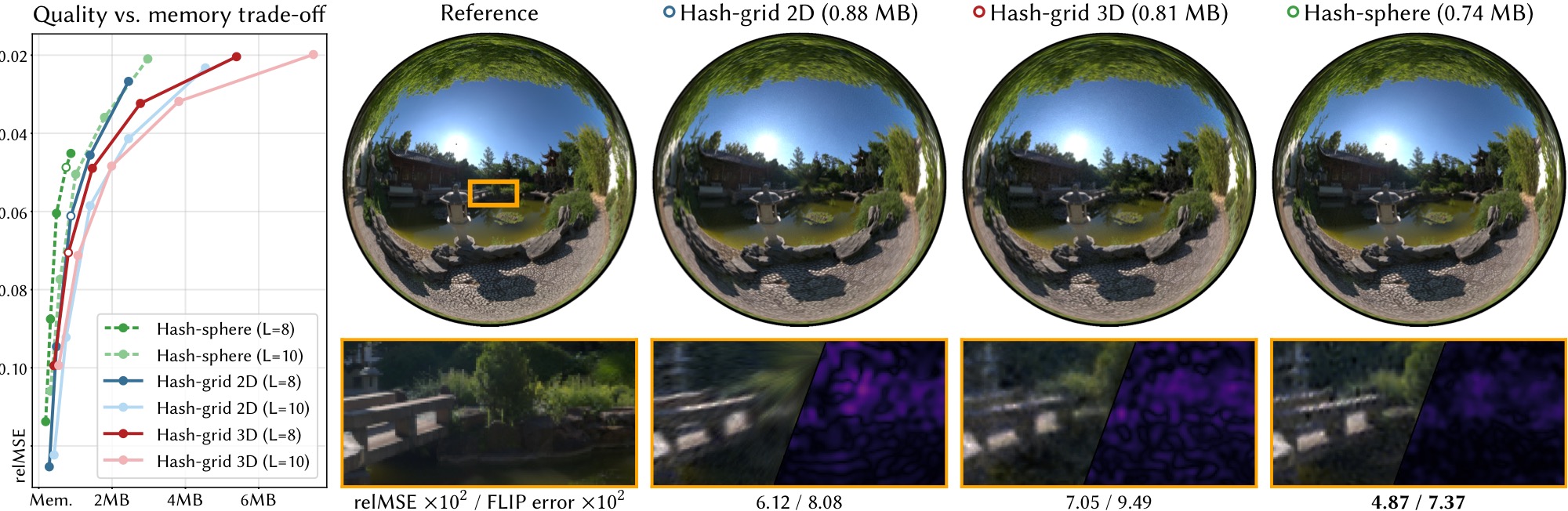}
    \\ \vspace{0.5cm}
    \includegraphics[width=\linewidth]{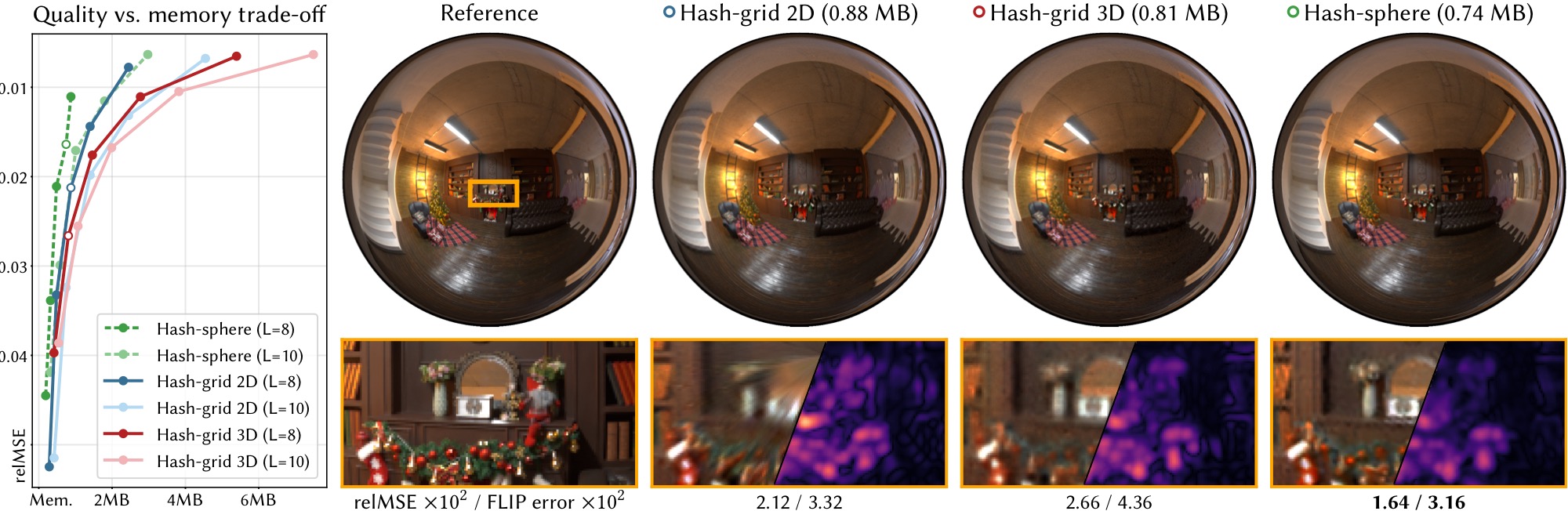}
    \\ \vspace{0.5cm}
    \includegraphics[width=\linewidth]{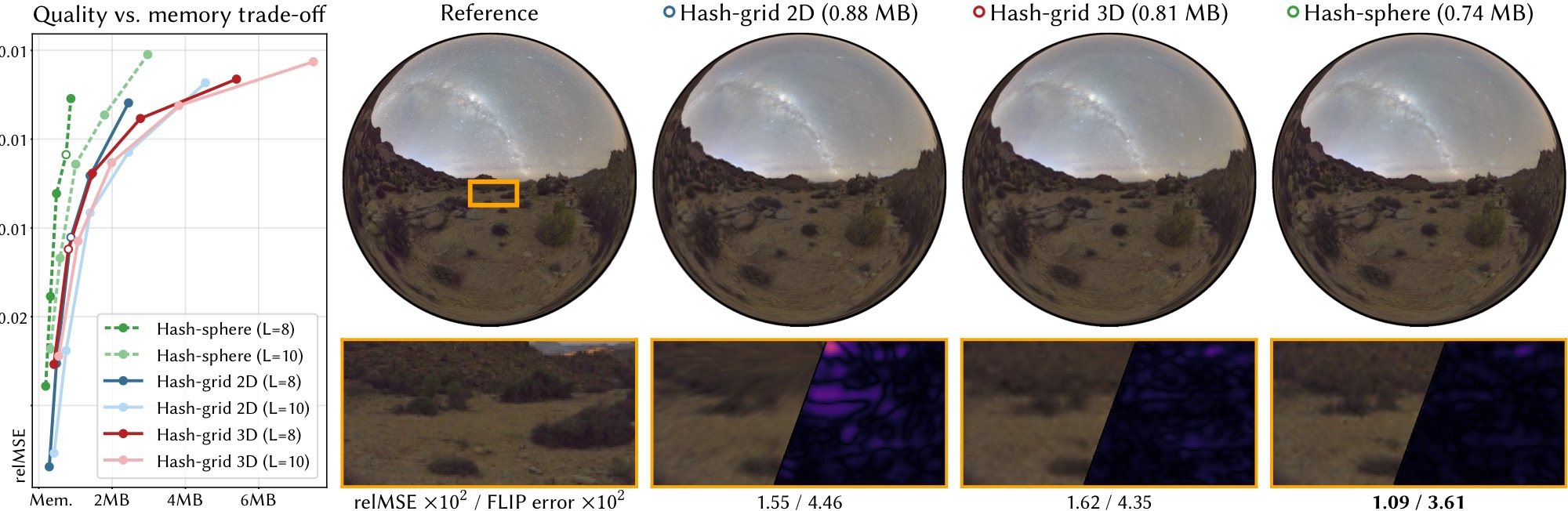}
    \caption{Additional quality vs. memory comparisons for HDR environment map compression. As in the main paper, the 2D hash-grid (polar parameterization) achieves comparable quality at mid-latitudes but suffers from severe distortions near the poles, while our \hashsphere\ provides consistent angular resolution across the entire sphere. For the hash-grid variants, we set the base resolution to 8 and per-level scaling to 2, all encodings use 2 features per level, testing configurations with various numbers of levels $L$ while varying the hashmap sizes.}
    \label{fig:additional_directional_encoding}
\end{figure*}

\Cref{fig:additional_directional_encoding} shows additional comparisons for the HDR environment map compression task described in the main paper. Across all tested environment maps, our \hashsphere\ consistently avoids the polar distortions exhibited by the 2D hash-grid while achieving comparable or better quality-versus-memory trade-offs.
\bibliographystyle{ACM-Reference-Format}
\bibliography{bibliography}